\documentclass{PoS}
\input epsf 
\newcommand{\beq}{\begin{equation}}
\newcommand{\eeq}{\end{equation}}
\newcommand{\bea}{\begin{eqnarray}}
\newcommand{\eea}{\end{eqnarray}}
\newcommand{\D}{\displaystyle}

\newcommand{\ssc}{\scriptscriptstyle}

\newcommand{\vev}[1]{\Big\langle #1 \Big\rangle}
\newcommand{\bpsi}{\bar{\psi}}

\title{Fermion RG blocking transformations and IR structure}

\ShortTitle{Fermion RG and IR structure}

\author{X. Cheng\\
        Department of Physics and Astronomy, University of California, Los Angeles\\
        Los Angeles, CA 90095, USA\\
        E-mail: \email{darktree@physics.ucla.edu}}

\author{\speaker{E.T. Tomboulis}\\
        Department of Physics and Astronomy, 
        University of California, Los Angeles\\
        Los Angeles, CA 90095, USA\\
        E-mail: \email{tomboulis@physics.ucla.edu}}

\abstract{We explore fermion RG block-spinning transformations on the lattice with the aim of studying the IR structure of gauge theories and, in particular, the existence of IR fixed points  for varying fermion content. In the case of light fermions the main concern and difficulty is ensuring locality of any adopted blocking scheme.  We discuss the problem of constructing a local blocked fermion action in the background of arbitrary gauge fields. We then discuss the carrying out of accompanying gauge field blocking. In the presence of the blocked fermions 
implementation of MCRG is not straightforward. By adopting judicious approximations we arrive at  an easily implementable approximate RG recursion scheme that allows quick, inexpensive estimates of the location of conformal windows for various groups and fermion representations.  We apply this scheme 
to locate the conformal windows in the case of SU(2) and SU(3) gauge groups. Some of the reasons for the apparent efficacy of this and similar decimation schemes are discussed.}

\FullConference{International Workshop on QCD Green's Functions, Confinement and Phenomenology,\\
		September 05-09, 2011\\
		Trento Italy}

\begin{document}

\section{Introduction}

The dependence of the phase structure of  
non-Abelian gauge theories on their fermion content is a question of obvious field theoretic and physical interest. In view of anticipated discoveries at the LHC this question has recently come to the fore with some urgency. For sufficiently small number of fermions there is UV asymptotic freedom (AF) and IR confinement. It is also known that 
for sufficiently large number of fermions asymptotic freedom and confinement are lost. 
It is generally believed that there is an intermediate range with an UV AF 
fixed point but no confinement; instead there should be an IR fixed point 
resulting in conformal IR dynamics. 
The dependence of these `conformal windows' on the 
gauge group, the number of fermion multiplets and their representation, as well as the structure of the 
associated IR fixed points, i.e. 
their relevant directions and anomalous dimensions,  are questions of 
central importance for informed model building "beyond the standard
 model".  For example, the viability of current walking technicolor models depends on such detailed 
knowledge.

Several approaches have been employed over the last few years in order to investigate these issues 
(see \cite{DelD} - \cite{Latproc} for review and references). 
One popular method is to adopt some definition of a running coupling, e.g. the 
`Schroedinger functional' definition.  Since any definition of a `coupling' outside the perturbative regime is necessarily arbitrary, any such choice is scheme-dependent. Caution should then be exercised in interpreting results. 
Another approach is to compute the spectrum and/or some judiciously chosen  physical observable and track their deformation as a function of the number of fermions and their masses. Such spectral studies should, in principle, yield a clear outcome, but they are very expensive to carry out in practice to the degree of accuracy required for unambiguous results. Furthermore, it is not always clear what the expected behavior should be. 

A direct method for getting at the phase structure 
would be implementation of the Wilsonian Renormalization Group (RG). Successive RG blocking 
transformations determine the effective action RG flow and its fixed points. 
Conceptually, this is the most straightforward and unambiguous approach. 
Various RG-based studies have been carried out recently for the most part using the two-lattice matching MCRG method \cite{AH1} - \cite{CDGK2}. 
There are, however, special challenges that must be met when attempting to implement RG blockings and MCRG methods in the presence of light fermions. 

In the following we will examine some of the issues involved in performing RG transformations with 
fermions. We will examine the question of preserving locality of the resulting action after 
an RG blocking step of light fermion degrees of freedom.
For full RG implementation such a step must of course be accompanied by a gauge field blocking step. Again, how to do this in practice in the presence of fermions, in particular how to implement  MCRG, is not straightforward. To get some insight we will adopt some judicious approximations that lead to an easily implementable approximate RG recursion scheme \cite{CT}. This scheme, which proves rather effective, is then applied to locating the conformal windows in the case of the $SU(2)$ and $SU(3)$ groups. Some or the reasons for its apparent effectiveness are discussed in the last section.

\section{Fermion RG blocking}
In devising RG blocking transformations for light fermions the main concern is maintaining reasonable locality of the blocked action after each blocking step. 
This is a question of interest in it's own right apart from issues of practical 
implementation. 
For {\it free} fermions a blocked action can be defined along the lines below (eq. (\ref{PF1})-(\ref{Q1})) that can be shown to maintain locality \cite{Betal}. 
This has been used to argue \cite{Sh} the locality of determinant rooting for free fermions.
This locality demonstration, however, cannot be extended in the presence of an 
arbitrary gauge field background. 

To begin exploring this issue consider the Wilson operator in background gauge field $U$: 
\beq
D_0(U) = 1 - \kappa M[U] ,  \label{act1}
\eeq
where the only non-vanishing elements of matrix M are:
\beq
M_{n\ (n+\hat{\mu})}[U]  =  (1-\gamma_\mu)\,U_\mu(n) \; , \qquad 
M_{n\ (n-\hat{\mu})}[U]  =  (1+\gamma_\mu)\,U_\mu^\dagger(n-\hat{\mu})
\, .\label{act1a}\eeq
A fermion RG blocking step specifies a transition from the fermion degrees of freedom 
on the original lattice $\Lambda$ to `thinned-out' degrees of freedom  on a lattice $\Lambda^\prime$. 
With $\Lambda$ a lattice of spacing $a$, let 
$\Lambda^{(1)}$\ denote the lattice  of spacing $2a$. We take $\Lambda^\prime = \Lambda^{(1)}\cup S$, where $S$ is a subset of $\Lambda$ to be specified later.

We introduce variables $\Psi^{\ssc(1)}$ on \quad  $\Lambda^\prime = \Lambda^{(1)}\cup S$ \quad by 
\bea 
& & \int D\bar{\psi}\,D\psi \; \exp(-\bar{\psi} D_0 \psi ) \nonumber \\
& = &\!\! \int D\bar{\psi}\,D\psi D\bar{\Psi}^{\ssc (1)}\,D\Psi^{\ssc (1)} \;
\exp\left\{-\bar{\psi} D_0 \psi - \alpha \left(\bar{\Psi}^{\ssc(1)} - \bar{\psi}Q^\dagger\right) 
\left(\Psi^{\ssc(1)} - 
Q\psi\right) \right\} \,. \label{PF1}
\eea
The rectangular matrices $Q\,: \Lambda \to \Lambda^\prime$ are given by 
\bea 
Q_{nm}\psi_m & =&  \psi_n +\zeta \sum_{\mu=1}^d M_{n\,(n+\hat{\mu})}[U] \psi_{(n+\hat{\mu})} 
+ \zeta \sum_{\mu=1}^d M_{n\,(n-\hat{\mu})}[U] \psi_{(n-\hat{\mu})}  \label{Q1} \\
& \equiv & \left( \delta_{nm} + \zeta \tilde{Q}_{nm}\right) \psi_m   \, .  \label{Q2}
\eea
Note that $\alpha \to \infty$ gives a $\delta$-function definition of the $\Psi^{\ssc(1)}$'s, i.e. 
$\Psi_n^{\ssc(1)}=( Q\psi )_n $. 
Integrating out the original variables one has on the new lattice $\Lambda^\prime$:
\bea 
& & 
\int_{\Lambda} D\bar{\psi}\,D\psi \; \exp(-\bar{\psi} D_0 \psi ) \nonumber \\
& = &   {\rm Det} G^{-1}\, \int_{\Lambda^{\prime}}  D\bar{\Psi}^{\ssc (1)}\,D\Psi^{\ssc (1)} \; 
\exp\left\{ - \alpha \bar{\Psi}^{\ssc(1)} \Psi^{\ssc (1)} 
+ \alpha^2\, \bar{\Psi}^{\ssc(1)} Q G Q^\dagger\Psi^{\ssc(1)}  \right\} \label{PF2}
\eea
with 
\beq
G^{-1} = \left[\, D_0 + \alpha \,Q^\dagger Q\,\right] \nonumber \,.
\eeq
In general, with light fermions, $G$ is a very non-local propagator.

\begin{figure}
\begin{center}
\includegraphics[width=9cm]{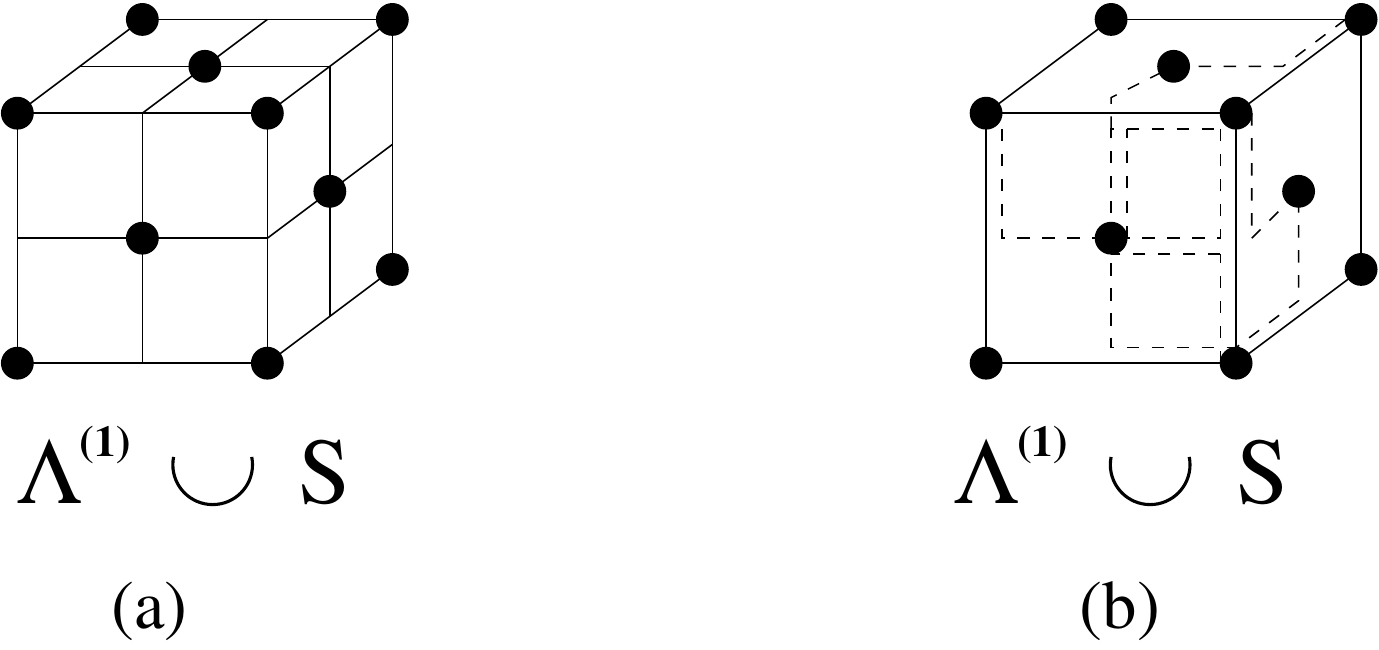}
\caption{(a) The set $S$, as a subset of $\Lambda$,  is the set of sites reached from the interior site of each  2x2 hypercube by displacement by $a$ in each lattice direction; (b) fermions at sites in $S$ interact with fermions on $\Lambda^{(1)}$ and among themselves by `fat links' consisting of all 
possible `staples' of length $2a$, a subset of which is indicated here by broken lines. \label{F1}}
\end{center}
\end{figure}

We now choose the set $S$ and the decimation parameter $\zeta$ in (\ref{Q1}) so that in the action 
\[ -\bar{\psi} D_0 \psi - \alpha \left(\bar{\Psi}^{\ssc(1)} - \bar{\psi}Q^\dagger\right) 
\left(\Psi^{\ssc(1)} - Q\psi\right) \] 
in (\ref{PF1}) one cancels the nontrivial part of $D_0$: 
\beq
 \bar{\psi}\left[ \kappa M - \alpha\zeta ( \tilde{Q} + \tilde{Q}^\dagger ) \right]\psi = 0 \label{actcancel} \, . 
 \eeq 
This is accomplished by setting $\zeta = \kappa/\alpha$, and taking the set $S$ as shown in Fig. 
\ref{F1}(a), i.e. $S$, viewed as a subset of $\Lambda$, is the set of the interior sites of the boundary plaquettes of the 2x2-hypercubes  that 
are the elementary d-cells of $\Lambda^{(1)}$. 
The propagator $G$ for the the original variables $\psi$ in (\ref{PF1}) is then given by 
\beq
G^{-1} = 1(\alpha) + {\kappa^2 \over \alpha}  \tilde{Q}^\dagger \tilde{Q}  > 0 \;,
\qquad  1(\alpha)_{nm} =  \left\{\begin{array}{c  l}            
(1+\alpha)\delta_{nm} & n,m \in \Lambda^\prime \\
\delta_{nm} & n,m \in \Lambda\setminus \Lambda^\prime
\end{array} \right. \; ,\label{prop}
\eeq
and the action in (\ref{PF2}) from the integration over the original $\psi$ becomes 
\beq
\bar{\Psi}^{\ssc(1)}\left[ {\alpha\over 1+\alpha} - \kappa^2 \tilde{Q} G \tilde{Q}^\dagger \right]\Psi^{\ssc(1)} \, .  \label{act2}
\eeq
(\ref{act2}) is manifestly local on $\Lambda^\prime = \Lambda^{(1)}\cup S$. 
In particular, note that from (\ref{prop}), taking $\alpha$ large, interactions induced by 
$G$ may be expanded in a hopping expansion in ${\kappa^2\over \alpha}$ and resummed.  
In fact, in the limit $\alpha\to \infty$, i.e. for the $\delta$-function definition of the blocked variables (cf. above), $G$ becomes ultralocal, and in fact $U$-independent, and the action goes to 
\beq
 \bar{\Psi}^{\ssc(1)}\left[ 1 - \kappa^2 \tilde{Q} \tilde{Q}^\dagger \right]\Psi^{\ssc(1)} \, .
 \label{act3}
 \eeq
After this blocking step we end up with fermions on the boundary of the blocked cube, i.e. a form of  {\it exact}  `potential moving' fermion blocking albeit on a non-hypercubic lattice. The fermion action is still bilinear in the blocked fermion fields which 
interact via `fat' gauge field bonds (Fig. \ref{F1}(b)). The gauge field bond variables are yet to 
be blocked.

It is important to note that if one starts integrating out gauge field bonds
inside each blocked hypercube, four-fermion and higher-fermion interactions will in general be induced 
(but of course at the scale of the blocked cube). 
Ultralocality of $G$ in this connection completely avoids potential difficulties 
from the ${\rm Det} G^{-1}$ factor in (\ref{PF2}) in performing any such accompanying gauge field blocking. 

We have devised a fermion blocking step in a general gauge field background that results in a manifestly local action. How to perform successive fermion blocking 
steps in a fixed gauge field background under the requirement that the action be manifestly local at the scale of each step is an interesting question in its own right. The following is a sufficient condition: for the n-th step implement the above blocking procedure integrating out the fermions on the sites satisfying the conditions:
\bea
x_i & = & 2^{n-1}m_i \;,  \qquad \qquad m_i \in Z, \quad i=1, \ldots, d \;,  \nonumber \\
\sum_i x_i & = & 2^n l + 2^{n-1} \;, \qquad \quad \; \;l\in Z \;.
\eea
Then the n-th step blocked fermions interact again via bilinear interactions through gauge field fat links of maximum length $2^n$ (in units of $a$). One can actually do better if one allows maximum 
length $2^{n+1}$ and performs a two-stage blocking of fermions interior to each blocked hypercube. 
The general lesson, though, is that one cannot achieve a {\it manifestly} local blocked fermion
action residing on a purely hypercubic lattice, i.e. without some non-empty analog of the set $S$ 
above (cf. Fig. \ref{F1}). This seems to accord with previous studies of fermion actions \cite{DeG}.

\section{Full RG blocking} 
For the physical questions we are here interested in, i.e. determination of the RG flow and its fixed points, RG blockings of both fermions and gauge fields must of course be implemented.
A local blocking step for fermions such as above must then be accompanied by a gauge field blocking step. 
It is of course possible to define a variety of gauge field blocking transformations, and several 
instances have been employed the literature, e.g. Swendsen, DSB or HYP-type transformations. 
Having defined the blocked gauge field variables, however, actually implementing the gauge blocking can generally be done only numerically. In the case of pure gauge theories, MCRG provides a tried method. Generally, given a starting action one generates a set of configurations by MC which are  blocked. An effective action on the blocked lattice is then determined by measuring its coupling on the blocked ensemble. Effective actions that reproduce expectations of observables over a wide range of scales can in principle be so constructed \cite{T1mcrg} - \cite{T2mcrg}. This procedures may be repeated over several blocking steps. If one is not interested in the effective action flow but only in observables, the two-lattice matching method \cite{AH3} may be used. Here one generates two ensembles of configurations, starting from two different 
couplings in the original action; a matching procedure is then applied to them by measuring observables on them.

In the presence of fermions, however, it is not clear exactly how such MCRG procedures are to be applied. 
One cannot generate a set of configurations to be blocked without first 
integrating out the fermions completely. One may then apply MCRG to this ensemble. In particular 
the two-lattice matching method has been applied in this manner in \cite{AH1} - \cite{CDGK2}.  
Eliminating the fermions of course generates a very non-local action for the gauge fields due to the resulting fermion determinant in the gauge field background, and naive universality notions do not necessarily hold. This raises some issues. 
It is not immediately apparent that measurements on repeatedly blocked configurations generated from this non-local action are equivalent to measurements using the local effective action in terms of the blocked fermion and gauge variables resulting from local blocking steps. In particular, it is not clear that for fermionic observables naively written in terms of simple fermion degrees of freedom on the blocked lattice sites some kind of universality guarantees equivalence to measurements with the local blocked action and observables in terms of blocked fermion degrees of freedom. Devising ways to test assumed equivalence of procedures, 
or designing new MCRG procedures for full fermion-plus-gauge field RG blocking still remain 
a largely unexplored area. 

Here, in order to get some insight, we opt for the modest goal of devising and trying out an approximate decimation scheme \cite{CT} that can easily be explicitly carried out.

\section{Approximate RG scheme} 

To obtain such a scheme we proceed as follows. 
Eliminate first the fermions on $S$ and compensate by adjusting the hopping parameter 
of the remaining fermions residing on the hypercubic lattice - this may be viewed as a typical fermion potential moving approximation to get a purely hypercubic action with `renormalized' gauge interactions (Fig. \ref{F2}). The renormalized hopping parameter is determined by requiring 
that some long-distance gauge-invariant fermion correlation function remains invariant under the fermion blocking step. Thus, taking the 2-point function, $\bpsi_n U[C_{nm}]\psi_m$, (other choices, such as a meson-meson correlator, will do equally well) this 
requirement translates into the exact relation: 
\begin{figure}
\includegraphics[width=7cm]{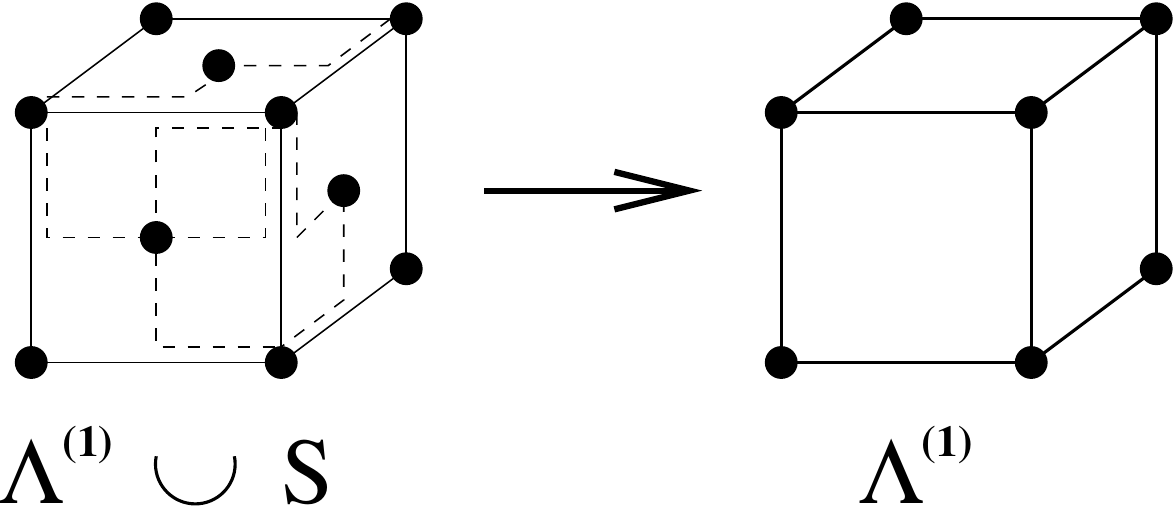} \hfill \includegraphics[width=0.35\textwidth]{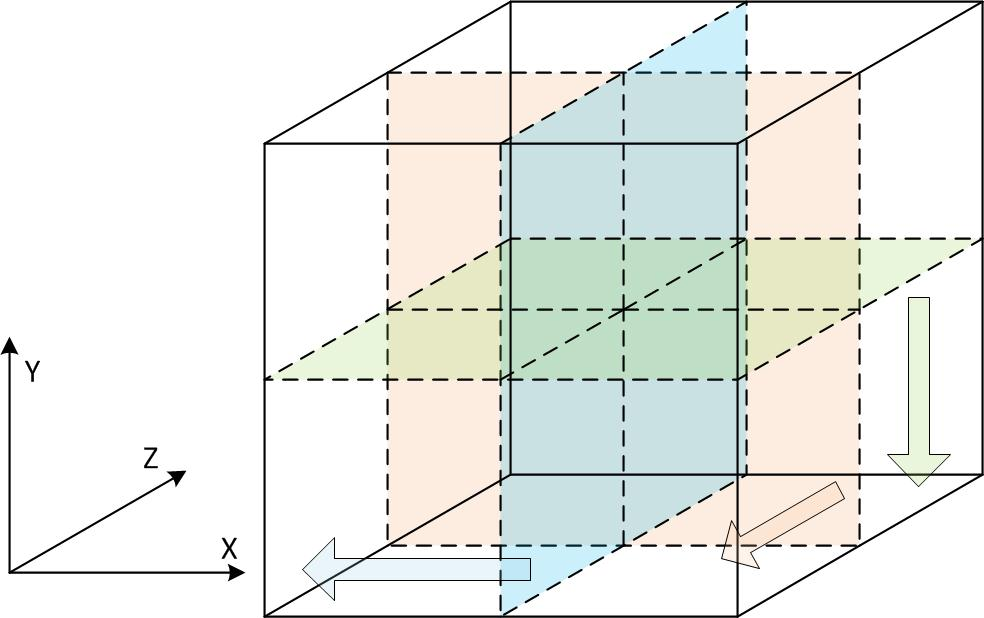}

\quad \qquad \qquad  \qquad (a)\hfill  (b)\qquad  \qquad \qquad\quad
\caption{Approximations: (a) Merge the fermions on $S$ from the exact blocking with the fermions on the hypercube 
edges while appropriately `renormalizing'   the hopping parameter of the latter, cf. text;
(b) standard potential moving approximation for gauge field blocking: move all interior plaquettes 
symmetrically to the boundary with appropriate weights. \label{F2}}
\end{figure} 
\beq 
\kappa(2a) =  D(\kappa(a)) \kappa(a)^2   \label{hop1}\, ,
\eeq
where 
\beq 
D(\kappa(a)) =  { \D \sum_{\mu,\nu} \vev{ \bar{\psi}_n (1-\gamma_\mu) (1-\gamma_\nu) \,
U[C_{nm}] U_\mu(n) U_\nu(n+\hat{\mu}) \psi_m} \over \D
\sum_\mu \vev{ \bar{\psi}_n (1-\gamma_\mu)  \,
U[C_{nm}] U_\mu(n) U_\mu(n+\hat{\mu}) \psi_m} } \; .\label{hop2}
\eeq 
(\ref{hop1})-(\ref{hop2}) provide a recursion relation relating the hopping parameters at lattice 
spacings $a$ and $2a$, which may, in particular, be solved to obtain the critical hopping parameter. 
One may then proceed to carry out the gauge field blocking step by whatever method is chosen. 

The above approximation maintains bilinear fermionic interactions at each 
successive blocking step. This suggests the following considerations. 
Assume one implements some RG blocking scheme. Denoting by $U^{(n)}$, $\bpsi^{(n)}$, 
$\psi^{(n)}$ the blocked variables at the $n$-th step, then, quire generally, one has 
\bea & &  
\int_{\Lambda^{(n)}} DU^{(n)} D\bar{\psi}^{(n)}  D\psi^{(n)}  \; \exp(- S_G^{(n)} - S_F^{(n)})    \nonumber \\
& =&  \int_{\Lambda^{(n+1)}} DU^{(n+1)} D\bar{\psi}^{(n+1)}  D\psi^{(n+1)}  \; \exp(- S_G^{(n+1)} - S_F^{(n+1)})   \;.    \label{RGrel1}
\eea 
If, furthermore, as after the above approximation, $S_F^{(n)}$ and $S_F^{(n+1)}$ maintain the same bilinear form on $\Lambda^{(n)}$ and $\Lambda^{(n+1)}$, respectively, then 
\beq
\exp (- S_G^{(n+1)}) = \int_{\Lambda^{(n)}} \; DU^{(n)} \, F(U^{(n)},U^{(n+1)})\;
 \left[{{\rm Det}\, S_F^{(n)} \over {\rm Det}\, S_F^{(n+1)} }\right]
\, \exp(- S_G^{(n)})  \, ,\label{RGrel2} 
\eeq 
with $F(U^{(n)},U^{(n+1)})$ specifying the blocking relation between $U^{(n)}$ and $U^{(n+1)}$'s.  
The ratio of determinants in (\ref{RGrel2}) can now be expected to be reasonably local. 
Indeed,  major part of non-locality is manifestly cancelled in the ratio. This can be seen from the formal expansion of the determinants in terms of gauge field loops:   
\beq 
{\rm Det}\, S_F^{(n)} = \prod_{C} {\rm det}\, ( 1- \kappa^{|C|} M[C]) \;, \label{detlexp}
\eeq
where $M[C]$ denotes the path ordered product along the loop $C$ of the matrices $M(U)$ defining the fermion bilinear operator in $S_F^{(n)}$, the product in (\ref{detlexp}) being over all loops. 
All loops in the expansion of $\ln {\rm  Det}S_F^{(n+1)}$ then are contained in that of ${\rm  Det}S_F^{(n)}$.  
The main contribution to the ratio comes in fact from loops inside each blocked hypercube. 
This suggest that one may approximate it by just such a set. 

\subsection{Simplest RG implementation model} 
These considerations lead us to an approximate implementation of RG decimations of the full system of fermions and gauge fields given by the following recipe.

After a fermion blocking step, apply the approximation depicted in Fig. \ref{F2}(a) combined with 
(\ref{hop1})-(\ref{hop2}).  Then the gauge field action on the spacing $2a$ lattice is given by 
(\ref{RGrel2}). As just explained, after the cancellation of the denominator by the numerator in the determinant ratio expressed in terms of gauge field loops, we can approximate the remainder by the contributions of loops interior to 
each blocked hypercube - in particular, by just the single plaquette loop weighted by a parameter $\eta$.
(One may of course consider approximations involving more interior loops, e.g. $2\times 2$ loops, and more decimation parameters.)  
It remains to specify and perform a gauge field blocking over each hypercube. This we do by 
standard potential moving  decimation: every interior plaquette is moved to the hypercube boundary with the appropriate weights $2^{d-2}$ (Fig. \ref{F2}(b)). Inserting these steps in (\ref{RGrel2}) 
one obtains 
\beq 
 \exp (-S_G^{(n+1)}[U[p^{(n+1)}]])  = \int  \prod_{b \in {\rm int}p^{(n+1)}}\!\! dU_b \;  
 {\rm det} (1- \kappa^4 M[p_0^{(n)}])^{2^d \eta} 
\prod_{p^{(n)} \in p^{(n+1)}} \left( \exp (-S_G^{(n)}[U[p^{(n)}]]) \right)^{2^{d-2}}    
\label{RGrel3}      
\eeq 
giving the plaquette action for the gauge field $S_G^{(n+1)}[U[p^{(n+1)}]]$ on lattice 
$\Lambda^{(n+1)}$ in terms of the theory on $\Lambda^{(n)}$. In (\ref{RGrel3}) 
$p^{(n)} \in p^{(n+1)}$ signifies the plaquettes $p^{(n)}$ in $\Lambda^{(n)}$ 
tiling plaquette $p^{(n+1)}$ in $\Lambda^{(n+1)}$, and the integrations are over those bonds $b\in \partial p^{(n)}$ that also belong to the interior of $p^{(n+1)}$. 
The plaquette $p_0^{(n)}$ is one of the tiling plaquettes (which one is immaterial by symmetry and translation invariance). Note that lattices $\Lambda^{(n)}$ and $\Lambda^{(n+1)}$ have lattice spacings differing by a scale factor of 2. 
To work out (\ref{RGrel3}) explicitly, all factors are expanded in group characters,  viz. 
\bea
\exp (-S_G^{(n)}[U[p^{(n)}]])& = & \left[ c_0(n) + \sum_j c_j(n) \chi(U[p^{(n)}])\right]  
\label{actexp1}\\
   & = & \exp \left[ \beta_0(n) + \sum_j \beta_j(n) \chi(U[p^{(n)}])\right] \;,\label{actexp2}
\eea
and similarly for $S_G^{(n+1)}[U[p^{(n+1)}]]$ and the determinant factor. Inserting in 
(\ref{RGrel3}), on the r.h.s. the various factor powers may be evaluated by the CG relations for combining  characters and the integrations may be carried out  exactly by using the character orthogonality relations. As a result 
one obtains explicit, albeit lengthy and ugly, recursion relations determining the expansion coefficients $c_j(n+1)$ at the $(n+1)$-th blocking step in terms of those at the $n$-the step. 
The procedure may then be iterated any number of times. It is important to note that the 
action at each step in general contains an infinite number of non-trivial characters, with corresponding couplings $\beta_j(n)$ (cf. (\ref{actexp2}). They will be generically generated just after a single blocking step even if one starts with a single representation action, e.g the Wilson fundamental representation action, on the original lattice.   

This RG recursion scheme depends only on two quantities: the parameter $\eta$, and, at each blocking step, the critical $\kappa(2^n a)$. 
We have applied it to $SU(2)$ and $SU(3)$. The results are shown in Table \ref{T1}. 
$N_{\rm L}$ and $N_{\rm U}$ denote the number of flavors at which an IR fixed point first appears and that at which asymptotic freedom is lost, respectively. 
The parameter $\eta$ was fixed by the conformal window upper bound value value $N_{\rm U}$ for $SU(2)$ fundamental representation fermions. This decimation parameter could in 
principle be tuned independently in each case, but it turns out that once set by the top entry in the second column, no further adjustment was found necessary in order to produce the rest of 
the entries with the correct $N_{\bf U}$ values. Having fixed the parameter $\eta$ the only uncertainties in practice are in obtaining the critical $\kappa$ values. 
They were obtained from estimates from (\ref{hop2}) and/or values from simulations reported in the literature \cite{CrK}.   
The resulting recursion relations for the character coefficients can then be run, very cheaply, to essentially arbitrary accuracy. It is noteworthy that such an approximate decimation scheme can already produce results such as shown in Table \ref{T1}. 
\begin{table} 
\begin{center}
\begin{tabular}{|l|l|l|}
\hline
  &$ N_{\rm L}$ & $N_{\rm U}$ \\
  \hline
SU(2) fund & 9 & 11 \\
SU(2) adjoint & 2 & 3\\
SU(3) fund & 11 - 12 &  17 \\
SU(3) sextet & 2 & 4 \\
\hline
\end{tabular}
\end{center}
\caption{Conformal window lower and upper bounds as obtained from the  
blocking recursion relations described in the text (section 4.1).\label{T1}}
\end{table} 
\section{Discussion - Outlook}
Elucidating the phase diagram of gauge theories with different fermion content is a fully nonperturbative problem that presents many challenges. All methods that have been 
tried so far have their practical or conceptual shortcomings. Here we have considered what should be conceptually the most straightforward approach, i.e. the flow of the system under successive 
RG blocking transformations.

We have first explored locality-preserving RG blocking schemes for light fermions. We saw that 
blockings maintaining locality in an arbitrary gauge field background can be devised, but they 
generically require non-hypercubic blocked fermion actions. 
Such schemes may prove useful in investigating a variety of interesting issues and applications, 
e.g. the `rooting' of fermion determinants; or improvements in the derivation of effective actions 
for exploring the QCD phase diagram via strong coupling and hopping expansions 
\cite{Phetal1} - \cite{Phetal2} combined with simulations.

Full (fermion plus gauge field) RG blocking would seem to provide the most direct and reliable way for exploring the  IR phase structure with varying fermion content. Carrying out the gauge field 
blocking explicitly in the presence of the blocked fermions, however, presents new challenges. 
In particular, implementation of MCRG procedures, which have been applied in the pure gauge theory case, is no longer straightforward for the reasons outlined above. Much work 
remains to be done to sort out these issues. 

At the same time judiciously chosen simplifications lead to approximate schemes which prove, as we saw, surprisingly effective. We found, in particular, that combining fermion blocking with gauge blocking within a potential moving approximation yields very reasonable results compared to  
other computationally much more expensive methods that have being tried. 
It is interesting to consider why this might be so. As it is well-known in the pure gauge case 
potential moving gives rather accurate beta functions at weak coupling (within $2\%$ for 
$SU(2)$). When fermions are coupled to the system, appropriate weighing of the fermion contribution by a decimation parameter (such as the parameter $\eta$ above) apparently can preserve this feature. 
This ensures that the upper limits of conformal windows as predicted by perturbation theory are correctly reproduced. It generally sets the flow `on the right path'. It is very important in this connection that this flow 
moves off into an infinite dimensional subspace of couplings (cf. (\ref{actexp2})) - we know 
from MCRG studies \cite{T1mcrg}, that including enough characters in the effective action is 
important for good behavior over a wider range of scales.  Moving in a sufficiently large, and preferably infinite dimensional, coupling space is crucial for evading lattice artifacts such as critical points and boundaries that can appear when projecting into a small finite-dimensional coupling constant subspace. 

If one is interested only in locating fixed points, rather than in detailed comparisons of the behavior of observables, following the RG flow provides a `robust' method in the following sense. As long as the 
flow manages to be within the right domain of attraction, it will lead to the corresponding fixed point; high accuracy in obtaining an effective action flow is not required. Approximate blocking schemes that incorporate the right features necessary to achieve this are then perfectly adequate for the purpose. This appears to indeed be the case for the scheme considered here. This minimal scheme, and other similarly constructed decimation schemes, are amenable to refinement by the inclusion of additional decimation parameters in more elaborate blocking definitions. It remains to be seen what improvements, if any, can be achieved by such refinements.

\medskip 
The authors would like to thank the organizers for the invitation to present this work and for 
organizing such a stimulating and enjoyable workshop. This research was partially supported by NSF PHY-0852438.

\end{document}